\documentclass[aps, prb, notitlepage, superscriptaddress, longbibliography, twocolumn, floatfix, 10pt]{revtex4-2}

\usepackage{xcolor}
\usepackage{grffile}
\usepackage{amsmath, amsthm, amssymb, bbold}
\usepackage[normalem]{ulem}
\usepackage{cancel}
\usepackage{bm}
\usepackage[nopatch=footnote]{microtype}
\usepackage{hyperref}
\usepackage{setspace}
\usepackage{grffile}
\hypersetup{colorlinks=true,linkcolor=blue,urlcolor=blue,citecolor=red}
\usepackage{amsmath}    % need for subequations
\usepackage{amssymb}
\usepackage{graphicx}   % need for figures
\usepackage{verbatim}   % useful for program listings
\usepackage{color}      % use if color is used in text
\usepackage[normalem]{ulem}
\usepackage{enumitem} 
\usepackage{dsfont} 
\usepackage{multirow}

\newcommand{\average}[1]{\langle #1\rangle}

\begin{document}

\title{Dynamic scaling and Family--Vicsek universality in the Hubbard model at infinite temperature}

\author{C\u at\u alin Pa\c scu Moca}
\email{mocap@uoradea.ro}
\affiliation{Department of Physics, University of Oradea,  410087, Oradea, Romania}
\affiliation{Department of Theoretical Physics, Institute of Physics, Budapest University of Technology and Economics, M\H{u}egyetem rkp.~3, H-1111 Budapest, Hungary}
\affiliation{MTA-BME Lend\"ulet "Momentum" Open Quantum Systems Research Group, Institute of Physics, Budapest University of Technology and Economics,
M\H uegyetem rkp. 3., H-1111, Budapest, Hungary}
\author{Doru Sticlet}
\affiliation{National Institute for R\&D of Isotopic and Molecular Technologies, 67-103 Donat, 400293 Cluj-Napoca, Romania}
\author{Bal\'azs D\'ora}
\affiliation{Department of Theoretical Physics, Institute of Physics, Budapest University of Technology and Economics, M\H{u}egyetem rkp.~3, H-1111 Budapest, Hungary}
\affiliation{MTA-BME Lend\"ulet "Momentum" Open Quantum Systems Research Group, Institute of Physics, Budapest University of Technology and Economics,
M\H uegyetem rkp. 3., H-1111, Budapest, Hungary}

\date{\today}
\begin{abstract}
We study Family--Vicsek scaling of charge, spin, and energy fluctuations in the one-dimensional Hubbard model at infinite temperature.
Using a quantum generating function approach, we compute time-dependent cumulants of transferred conserved quantities and analyze how the corresponding roughness depends on subsystem size and time.
We start by focusing on a single interface at half the chain and determine the transport exponents. Then we turn to fluctuations of a small finite interval and
study the Family--Vicsek universality of fluctuations over an extended timescale.
We find  that the long-time scaling behavior is controlled by integrability. In the free limit, charge, spin, and energy all display ballistic transport. 
In the interacting integrable Hubbard chain, charge and spin cross over to a KPZ scaling regime, while the energy sector remains ballistic. Once integrability is 
broken by a next-nearest-neighbor density interaction, the long-time dynamics becomes diffusive in all sectors. 
In every case we also observe a short-time microscopic regime with apparently universal ballistic growth before the hydrodynamic scaling window sets in.
The Family--Vicsek setup allows us to determine the growth, the saturation as well as the dynamical exponents.
\end{abstract}
\maketitle

\section{Introduction}
\label{sec:introduction}
The Family--Vicsek (FV) scaling~\cite{Vicsek1984,family1985scaling} is a universal framework originally introduced to
characterize the roughening of growing interfaces in nonequilibrium classical systems~\cite{hohenberg1977theory}.
Beyond surface growth, FV-type scaling ideas have found broad applicability in diverse settings, ranging from
fluid dynamics~\cite{dominguez2007scaling} to biological growth~\cite{family1991dynamics,barabasi1995fractal}.
In its simplest form, FV scaling asserts that the roughness $W(\ell,t)$ of fluctuations in a subsystem of
linear size $\ell$ follows the scaling law
\begin{equation}
W(\ell,t) \sim \ell^{\alpha}\, f\!\left(\frac{t}{\ell^{z}}\right),
\label{eq:FV_scaling}
\end{equation}
where $f(x\ll 1)\sim x^{\beta}$ and $f(x\gg 1)\sim \mathrm{const}$. The exponents satisfy $z=\alpha/\beta$, with
$z$ the dynamical exponent.

Recently, FV scaling has been discussed in quantum many-body settings~\cite{fujimoto2020family,fujimoto2021dynamical,fujimoto2022impact,glidden2021bidirectional,aditya2024family,luis2022unveiling,jin2022kardar,bhakuni2024dynamic},
where the role of the fluctuating height field can be played by an extensive conserved quantity in a subsystem.
In the one-dimensional Hubbard model~\cite{essler2005one}, the natural quantities to examine are subsystem-resolved charge, spin,
and energy observables. The corresponding roughness is defined from the growth of the second
cumulant of the transferred quantity.

While FV-type scaling has been established in a number of quantum models, much of the existing work has focused on
noninteracting systems or on interacting limits where additional analytical simplifications apply~\cite{fujimoto2020family,aditya2024family}.
In previous works~\cite{Moca.2026,Moca.2026two} we investigated the FV scaling of spin fluctuations in one-dimensional spin
chains, finding that the scaling behavior is controlled by integrability: integrable cases display
ballistic, KPZ, and diffusive scaling, while non-integrable cases always show diffusive scaling~\cite{Moca.2026two}.

The fermionic Hubbard model, on the other hand, is a paradigmatic model for correlated quantum dynamics and transport,
featuring intertwined charge and spin degrees of freedom and, in one dimension, integrability with an extensive set of
conserved quantities~\cite{essler2005one}. 

At low energies the one-dimensional Hubbard model provides a canonical realization of Luttinger-liquid phenomenology,
including charge--spin separation~\cite{Voit.1993,Voit_1995}, while at infinite temperature the same microscopic
conservation laws control fluctuation growth and hydrodynamic scaling windows.
At infinite temperature (or, more generally, at high temperatures compared to microscopic energy scales), transport and
fluctuation dynamics in the Hubbard chain provide an  arena to connect microscopic conservation laws, integrability,
and emergent hydrodynamics. The model is exactly solvable in one dimension~\cite{LiebWu,Shastry.86,Schlottmann.1997} and
its symmetry structure includes nontrivial spin and pseudospin generators~\cite{Grosse.1989}. In the nonequilibrium and
hydrodynamic descriptions of integrable systems, generalized hydrodynamics (GHD) provides a natural language to describe
ballistic and diffusive contributions to charge/spin/energy currents~\cite{Castro.2016,doyon.2023,Nozawa.2020,Nozawa.2021}.
At the same time, explicit high-temperature transport studies have established diffusive regimes in the Hubbard model in
appropriate symmetry sectors~\cite{Benenti,Prosen.2012} and have uncovered crossovers and superdiffusive windows tied to the
interplay of conserved quantities and non-Abelian symmetries~\cite{Fava.2020,ilievski2017microscopic,ilievski2021superuniversality,Bulchandani_2021}.
For broader context on finite-temperature transport in one-dimensional lattice models (including the Hubbard chain), see
the recent review~\cite{Bertini.2021}.
Recent work has connected these anomalous regimes to KPZ-type scaling in integrable quantum dynamics, including in the
Hubbard model itself~\cite{moca2023kardar,Ye.2022,Wei.2022}.

From the experimental side, cold-atom quantum simulators provide direct access to Hubbard dynamics and transport at
effective high temperatures and far from equilibrium~\cite{Schneider.2012,Cheneau2012}, thus motivating the development of theoretical tools to analyze fluctuation dynamics in this regime. 

In this work, we develop an FV scaling analysis for the one-dimensional Fermi--Hubbard model at \emph{infinite temperature}, where the dynamics is encoded in time-dependent
fluctuations. We analyze charge, spin, and energy on the same footing in order to compare how different conserved quantities probe the same underlying many-body dynamics.
To access the time-dependent fluctuations we use a quantum generating function (QGF) approach~\cite{Valli.2025,Moca.2026},
which yields cumulants of transferred conserved quantities without reconstructing the full distribution. In the noninteracting
limit $U=0$, we obtain analytical expressions and recover ballistic scaling. For finite interactions, charge and spin show KPZ scaling in the integrable chain, whereas the energy sector remains ballistic. Finally, upon
breaking integrability (here by adding a next-nearest-neighbor density interaction) the dynamics crosses over to
diffusion in all sectors.

The paper is organized as follows: In Sec.~\ref{sec:model} we introduce the model Hamiltonian and its symmetries,
including the integrability-breaking perturbation. In Sec.~\ref{sec:QGF} we present the QGF framework used to compute the relevant cumulants at infinite temperature. In Sec.~\ref{sec:dynexp}, we extract dynamical exponents from the growth of
half-chain moments with a single interface, profiting from the high quality numerical data. 
In Sec.~\ref{sec:fv_scaling} we analyze Family--Vicsek scaling collapses in the charge, spin, and energy sectors. Finally, in Sec.~\ref{sec:conclusions} we summarize our results and discuss possible extensions.

\section{Model Hamiltonian and symmetries}
\label{sec:model}

We study the one-dimensional Fermi--Hubbard model on a chain of $L$ sites,
\begin{gather}
H_{\rm Hubb}=-J\sum_{j=1}^{L-1}\sum_{\sigma=\uparrow,\downarrow}\left(c^{\dagger}_{j\sigma}c_{j+1,\sigma}+\mathrm{h.c.}\right) \nonumber\\
+U\sum_{j=1}^{L}\left( n_{j\uparrow}-{1\over 2}\right) \left(n_{j\downarrow}-{1\over2}\right),\label{eq:H_Hubbard}
\end{gather}
where $c^{(\dagger)}_{j\sigma}$ annihilates (creates) a fermion with spin $\sigma$ on site $j$ and
$n_j\equiv\sum_{\sigma}n_{j\sigma}$ with $n_{j\sigma}=c^{\dagger}_{j\sigma}c_{j\sigma}$, the local 
density operator. We measure energies in units of $J$ and set $J=1$ throughout.

The Hubbard Hamiltonian conserves total charge $N=\sum_j n_j$ and the total spin operators.
In the absence of a Zeeman field, it is invariant under global spin rotations, i.e., it has an $SU(2)$ spin symmetry
generated by
\begin{gather}
S^+ =\sum_{j} c^{\dagger}_{j\uparrow}c_{j\downarrow},\qquad
S^- = \sum_{j} c^{\dagger}_{j\downarrow}c_{j\uparrow},\nonumber\\
S^z=\frac{1}{2}\sum_j\left(n_{j\uparrow}-n_{j\downarrow}\right).
\label{eq:spin_SU2}
\end{gather}
This $SU(2)$ spin symmetry can be broken down to  $U(1)$ symmetry by an external magnetic field or a spin imbalance in the system
($\average{n_{\uparrow}} \neq \average{n_{\downarrow}}$), associated with the conservation of the $S_z$ component of the total spin.
The second non-Abelian symmetry of the model is the $SU_c(2)$ charge symmetry, reflecting the conservation of the $SU(2)$
$\eta$-operator with the components
\begin{gather}
	\eta^\dagger = \sum_{j=1}^{L} (-1)^j c^\dagger_{j\uparrow} c^\dagger_{j\downarrow},
	\quad \eta = \sum_{j=1}^{L} (-1)^j c_{j\downarrow} c_{j\uparrow}  \nonumber\\
	\eta_z =  \frac{1}{2}\sum_{j=1}^{L}\left(n_j-1\right).
\end{gather}
At half filling, the generators $(\eta^\dagger,\eta,\eta_z)$ close an $SU(2)$ algebra (often referred to as the
\emph{$\eta$-pairing} or pseudospin symmetry)~\cite{essler2005one,Moudgalya.2020}. This additional non-Abelian charge symmetry is special to half filling
and relies on the bipartite structure encoded by the staggered phase factor $(-1)^j$. Away from half filling, the pseudospin symmetry is reduced
to the Abelian $U(1)$ symmetry associated with conservation of the total particle number, with total charge $N=2\eta_z+L$.
The one-dimensional Hubbard model is integrable and admits a Bethe-ansatz solution~\cite{essler2005one,LiebWu,Shastry.86,Schlottmann.1997}.
Further Bethe-ansatz based analysis of wave functions and correlations can be found in, e.g., Ref.~\cite{Ogata.1990}.
To investigate how universality classes inferred from FV scaling evolve under integrability breaking, we add a
next-nearest-neighbor density interaction,
\begin{equation}
H_{\rm NNN}=V\sum_{j=1}^{L-2}(n_j-1)(n_{j+2}-1),\label{eq:H_NNN}
\end{equation}
and consider the non-integrable extended model
\begin{equation}
H=H_{\rm Hubb}+H_{\rm NNN}.
\label{eq:H_total}
\end{equation}
This perturbation preserves charge conservation and the global spin $SU(2)$ symmetry, but generically breaks
integrability (and, at half filling, also reduces the enlarged charge symmetry to $U(1)$), providing a controlled route
to contrast integrable and non-integrable transport regimes within the same FV scaling analysis~\cite{Surace.2023,Schagrin.2021,van1994extended,van1994extended2}.
Related symmetry-based reductions of the Hubbard Hilbert space and representation-theoretic perspectives are discussed in
Ref.~\cite{JAKUBCZYK2020293}.

\section{Quantum Generating Function approach}
\label{sec:QGF}

To access the full time-dependent fluctuations entering the FV scaling analysis, we use a quantum generating-function (QGF)
formulation which yields cumulants of extensive observables without reconstructing the full distribution~\cite{Moca.2026,SM}.
The MPO implementation we use naturally interfaces with non-Abelian symmetry resolution in Hubbard-type settings~\cite{Moca.2022}.
We work with a segment of length $\ell$ and focus on the \emph{occupation} (charge) operator
\begin{equation}
N_\ell \equiv \sum_{j\in\mathrm{seg}(\ell)}\sum_{\sigma=\uparrow,\downarrow} n_{j\sigma},
\end{equation}
and, in parallel, on the spin operator $S^z_\ell$. The two sectors are treated identically below,
so we introduce a generic segment observable $Q_\ell\in\{N_\ell,\,S^z_\ell\}$.

We consider unitary time evolution under the (closed) Hubbard Hamiltonian $H$ and an \emph{infinite-temperature} initial state.
We take the fully mixed density matrix,
\begin{equation}
\rho_{\infty}\equiv \frac{\mathds{1}}{4^L},
\end{equation}
which corresponds to half filling and zero magnetization on average; the formalism extends straightforwardly to
infinite-temperature states with fixed global $N$ and $S^z$ (projected sectors) or to grand-canonical constraints. The transferred variable associated with $N_\ell$ is
\begin{equation}
\Gamma_N(t)\equiv N_\ell(t)-N_\ell(0),
\label{eq:Gamma_N_def}
\end{equation}
and its second cumulant defines the roughness
\begin{equation}
W_N(\ell,t)=\sqrt{\kappa_2^{(N)}(\ell,t)}.
\label{eq:roughness_W}
\end{equation}
The central object is the twist operator
\begin{equation}
R_\ell^{(N)}(\lambda)\equiv e^{i\lambda N_\ell},
\label{eq:twist_operator}
\end{equation}
defined in terms of a (generally complex) counting field $\lambda$.
The unitary QGF (characteristic function) is defined as
\begin{align}
G_{\ell,N}^{(H)}(\lambda,t)
&\equiv \mathrm{Tr}\!\left[ R_\ell^{(N)}(\lambda)\, U(t)\, R_\ell^{(N)\dagger}(\lambda)\,\rho_{\infty}\, U^\dagger(t)\right],
\label{eq:QGF_unitary}
\end{align}
where $U(t)\equiv e^{-iHt}$.
Formally, $G_{\ell,N}^{(H)}(\lambda,t)$ is the characteristic function of $\Gamma_N(t)$ for the underlying two-time protocol,
and the corresponding distribution can be obtained by Fourier transform. In practice, we extract cumulants directly,
\begin{equation}
\kappa_n^{(N)}(\ell,t)=(-i)^n\,\partial_\lambda^n\left.\ln G_{\ell,N}^{(H)}(\lambda,t)\right|_{\lambda=0}.
\label{eq:cumulants_def}
\end{equation}
For sufficiently small $|\lambda|$, one may expand
\begin{equation}
G_{\ell,N}^{(H)}(\lambda,t)=1-\frac{\lambda^2}{2}\,\mu_2^{(N)}(\ell,t)+\mathcal{O}(\lambda^4),
\end{equation}
where $\mu_2^{(N)}(\ell,t)=\langle \Gamma_N(t)^2\rangle$ provided odd moments vanish by symmetry.
This yields the second moment (and cumulant) in the small-$\lambda$ limit,
\begin{equation}
\kappa_2^{(N)}(\ell,t)=\mu_2^{(N)}(\ell,t)\simeq \frac{2}{\lambda^2}\big(1-\mathrm{Re}\,G_{\ell,N}^{(H)}(\lambda,t)\big)+\mathcal{O}(\lambda^2).
\label{eq:kappa2_small_lambda}
\end{equation}
To suppress the leading $\mathcal{O}(\lambda^2)$ truncation error, it is advantageous to evaluate the QGF at two phases and
combine them~\cite{Valli.2025}. For example, choosing $\lambda=r$ and $\lambda=i r$ (with $r\in\mathbb{R}$) cancels the
$\mathcal{O}(r^2)$  contribution,
\begin{equation}
\mu_2^{(N)}(\ell,t)\simeq \frac{G_{\ell,N}^{(H)}(i r,t)-G_{\ell,N}^{(H)}(r,t)}{r^2}+\mathcal{O}(r^4).
\label{eq:mu2_phase_combo}
\end{equation}
If the chosen infinite-temperature ensemble yields $\langle N_\ell\rangle\neq 0$ (e.g., away from half filling or at finite
magnetization), one may work with the centered operator $\delta N_\ell\equiv N_\ell-\langle N_\ell\rangle$ (replacing $N_\ell$
in $R_\ell^{(N)}$) or evaluate $\kappa_1^{(N)}$ and $\kappa_2^{(N)}$ separately from Eq.~\eqref{eq:cumulants_def}.
Equation~\eqref{eq:QGF_unitary} is particularly well suited to tensor-network numerics because it reduces fluctuation
physics to a small number of real-time evolutions with modified (``twisted'') initial conditions.
In the MPO language, we prepare the twisted initial operator
\begin{equation}
\rho_\lambda^{(Q)}(0)=R_\ell^{(Q)\dagger}(\lambda)\,\rho_{\infty},
\end{equation}
evolve it unitarily to $\rho_\lambda^{(Q)}(t)=U(t)\rho_\lambda^{(Q)}(0)U^\dagger(t)$ using 
the time evolving block decimation (TEBD) in Liouville space~\cite{Moca.2026two}, and finally
evaluate
\begin{equation}
G_{\ell,N}^{(H)}(\lambda,t)=\mathrm{Tr}\big[ R_\ell^{(N)}(\lambda)\,\rho_\lambda^{(N)}(t)\big]
\end{equation}
by contracting the corresponding MPO tensors.
Because $R_\ell^{(N)}(\lambda)$ is close to the identity for small $|\lambda|$ and is supported only on the segment, the
operator entanglement generated in $\rho_\lambda^{(N)}(t)$ remains comparable to that of the physical evolution from
$\rho_{\infty}$, so cumulants can be extracted at a cost similar to standard real-time propagation.

\section{dynamical exponents}

Here we study the growth of fluctuation across a single interface only by focusing on a half chain partitioning. This allows us to extract the growth exponent unambiguously from
numerics, almost devoid of finite size effects. 

\label{sec:dynexp}
\subsection{Spin and charge fluctuations across an interface}\label{sec:dynexp_interface}
We extract dynamical exponents from the growth of charge and spin fluctuations across a single interface.
Concretely, we consider a bipartition of the chain into left/right halves and define the half-chain observables
\begin{equation}
N_{L/2}\equiv \sum_{j=1}^{L/2}\sum_{\sigma=\uparrow,\downarrow} n_{j\sigma},\qquad
S^z_{L/2}\equiv \frac{1}{2}\sum_{j=1}^{L/2}\left(n_{j\uparrow}-n_{j\downarrow}\right),
\end{equation}
which correspond to the segment definition in Sec.~\ref{sec:QGF} at $\ell=L/2$.
The central quantity in this section is the corresponding second moment (equal to the second cumulant for the protocols we consider),
\begin{equation}
\kappa_2^{(Q)}(t)\equiv \big\langle\Gamma_Q(t)^2\big\rangle,\qquad Q\in\{N,\,S^z\},
\label{eq:kappa2_half_chain}
\end{equation}
evaluated at infinite temperature.
\begin{figure}
\includegraphics[width=\columnwidth]{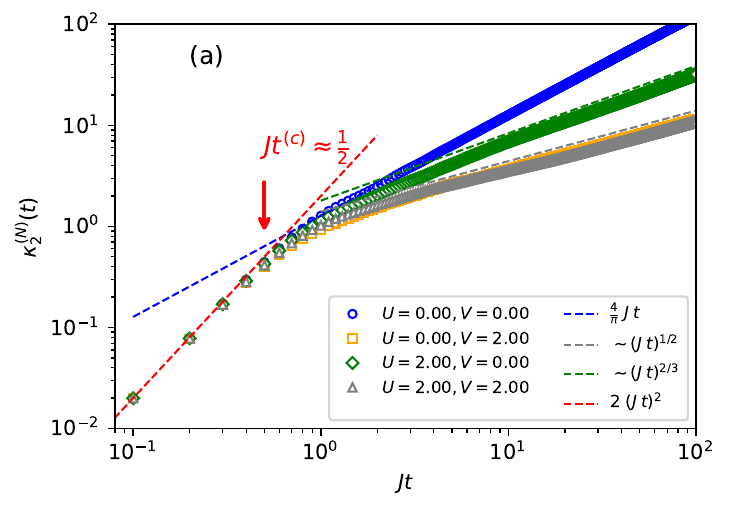}
\includegraphics[width=\columnwidth]{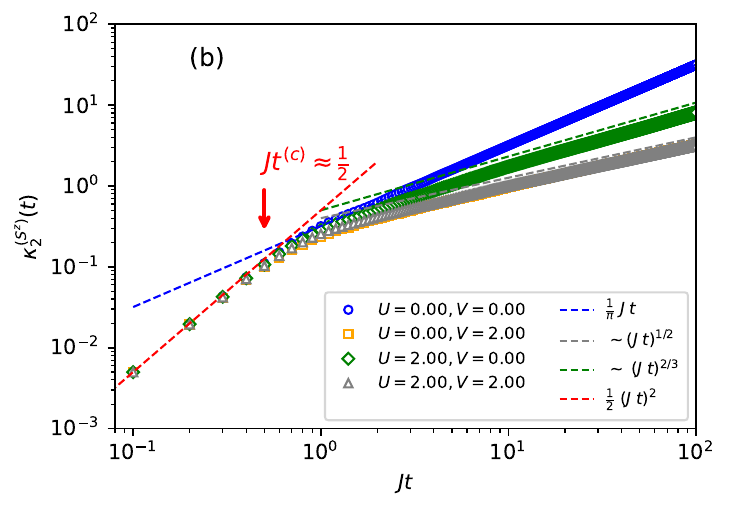}
\caption{Half-chain second moment $\kappa_2^{(Q)}(t)$ for charge (top) and spin (bottom)
at $U=0$ and $V=2$ (blue), $U=0$ and $V=2$ (orange), $U=2$ and $V=0$ (green), and $U=2$ and $V=2$ (gray). 
The dashed lines show the expected power-law growth with
exponents $1/z=1$ (ballistic), $2/3$ (KPZ), and $1/2$ (diffusive). 
For $U=0$ and $V=0$ the exact analytical results
from Eqs.~\eqref{eq:kappa2_U0_short_time} and \eqref{eq:kappa2_U0_asymptotics} are shown (dashed) and match the numerics.
The numerical data are obtained for $L=200$ sites and a small counting field $r=0.05$ in Eq.~\eqref{eq:mu2_phase_combo}.
The crossover time $t^{(c)}\sim 1/(2J)$ is indicated by the vertical red arrow.}
\label{fig:kappa2_half_chain}
\end{figure}
In the exactly solvable $U=0$ and $V=0$ limit, the half-chain second moment can be obtained analytically in the thermodynamic limit. Each spin
species $\sigma\in\{\uparrow,\downarrow\}$ evolves independently under the tight-binding Hamiltonian
\begin{equation}
H_{0,\sigma}=-J\sum_j(c^{\dagger}_{j\sigma}c_{j+1,\sigma}+\mathrm{h.c.}).
\label{eq:H0_sigma}
\end{equation}
The Heisenberg evolution is linear,
\begin{equation}
c_{j\sigma}(t)=\sum_m U_{jm}(t)\,c_{m\sigma},
\label{eq:U0_cj_evolution}
\end{equation}
with the translation-invariant kernel $U_{jm}(t)=i^{\,m-j}\,J_{m-j}(2Jt)$.
We consider infinite-temperature, translation-invariant ensembles which are diagonal in the occupation basis, so that
\begin{equation}
\langle c^{\dagger}_{m\sigma}c_{n\sigma'}\rangle= \langle c_{m\sigma}c^{\dagger}_{n\sigma'}\rangle =\frac{1}{2}\,\delta_{mn}\delta_{\sigma\sigma'}.
\end{equation}
Using Wick's theorem, one finds for equal-time densities 
\begin{equation}
\langle n_{i\sigma}(t)n_{j\sigma}\rangle-\bar n_{\sigma}^2
=\frac{1}{4}\,|U_{ij}(t)|^2.
\label{eq:U0_nn_kernel}
\end{equation}

One finds for each spin species $\sigma$,
\begin{equation}
\kappa_{2,\sigma}^{(0)}(t)=\big\langle\Gamma_{\sigma}(t)^2\big\rangle
=\frac{1}{2}\sum_{r=-\infty}^{\infty}|r|\,J_r^2(2Jt),
\label{eq:kappa2_half_U0_exact}
\end{equation}
where $J_r$ is the Bessel function of the first kind.
The charge and spin moments follow as
\begin{align}
\kappa_{2}^{(N)}(t)&=\kappa_{2,\uparrow}^{(0)}(t)+\kappa_{2,\downarrow}^{(0)}(t),\nonumber\\
\kappa_{2}^{(S^z)}(t)&=\frac{1}{4}\Big(\kappa_{2,\uparrow}^{(0)}(t)+\kappa_{2,\downarrow}^{(0)}(t)\Big).
\label{eq:kappa2_charge_spin_U0_exact}
\end{align}
At very short times one has $\sum_r |r|J_r^2(2Jt)=2J^2t^2+\mathcal{O}(t^4)$, and therefore 
$\kappa_2^{(Q)}(t)$ is quadratic in time, 
\begin{equation}
\kappa_2^{(N)}(t)\simeq 2J^2t^2,\quad \kappa_2^{(S^z)}(t)\simeq \frac{\kappa_2^{(N)}(t)}{4},\quad t\ll\frac{1}{2J}.\label{eq:kappa2_U0_short_time}
\end{equation}
In the transport (hydrodynamic) growth regime $1\ll 2Jt$ one may use the asymptotics
$\sum_{r=-\infty}^{\infty}|r|\,J_r^2(2Jt)\simeq \tfrac{4}{\pi}Jt$, which yields
\begin{equation}
\kappa_2^{(N)}(t)\simeq \frac{4}{\pi}Jt,\qquad \kappa_2^{(S^z)}(t)\simeq \frac{\kappa_2^{(N)}(t)}{4},
\quad t\gg\frac{1}{2J},
\label{eq:kappa2_U0_asymptotics}
\end{equation}
indicating a  ballistic linear growth in the long time limit,  $\kappa_2^{(Q)}(t)\propto t$ and thus $z=1$. Furthermore, the asymptotics indicate the existence of a crossover time scale of order 
\begin{equation}
t^{(c)}\sim \frac{1}{2J},
\end{equation}
separating the short-time quadratic growth from the long-time linear growth. 
This analytical result is supplemented by the numerical QGF evaluation. Within this approach the twist operator associated
with $Q\in\{N,S^z\}$ can be written as a product of local operators. For a half-chain cut one may write
\begin{gather}
	R_{L/2}^{(Q)}(\lambda) = \underbrace{e^{i\lambda Q_{1}}\otimes \dots \otimes e^{i\lambda Q_{L/2}}}_{1\le j\le L/2}\otimes
	\overbrace{\mathbb{1}\otimes\dots \otimes \mathbb{1} }^{L/2<j\le L}.
	\label{eq:twist_half_chain}
\end{gather}
In Fig.~\ref{fig:kappa2_half_chain}, we show the numerical results for $\kappa_2^{(Q)}(t)$. 
The symbols are obtained from the QGF. Some of the  dashed lines, corresponding to the $U=0, V=0$	 case, show the analytical expressions from
Eqs.~\eqref{eq:kappa2_U0_short_time} and \eqref{eq:kappa2_U0_asymptotics}, while others are guides to the eye for the expected power-law growth in the different regimes. The data are obtained for a chain of $L=200$ sites and a small counting field $r=0.05$ in Eq.~\eqref{eq:mu2_phase_combo}.
 The data match the analytical benchmark and
illustrate the crossover from $t^2$ to $t$ growth around $t^{(c)}\sim 1/(2J)$. Moreover, the early-time growth
$\kappa_2^{(Q)}(t)\propto t^2$ is independent of the interaction strength $U$ and of the integrability-breaking perturbation
$V$, while the late-time behavior is sensitive to both and can be summarized as
\begin{equation}
\kappa_2^{(Q)}(t)\propto \begin{cases}
t^2, & t\lesssim t^{(c)},\\
t, & t\gtrsim t^{(c)}, U=0, V=0,\\
t^{2/3}, & t\gtrsim t^{(c)}, U\neq 0, V=0,\\
t^{1/2}, & t\gtrsim t^{(c)}, V\neq 0.
\end{cases}
\label{eq:kappa2_scaling_z}
\end{equation}
At finite $U$ and $V=0$, the dynamical exponent is consistent with KPZ scaling, $z=3/2$, while the
non-integrable case with $V\neq 0$ shows diffusive scaling with $z=2$, irrespective of the charge/spin sector
and for the parameters we studied.

\subsection{Energy dynamical exponents}\label{sec:energy_dynexp}

In parallel with the charge and spin sectors, we consider the energy sector, which is also conserved under the
full Hubbard dynamics.
In analogy with Sec.~\ref{sec:dynexp_interface}, we define the half-chain energy observable. For the Hubbard Hamiltonian
it is natural to associate an energy density with each bond. 
The local bond energy is
\begin{align}
h_j &\equiv -J\sum_{\sigma=\uparrow,\downarrow}\!\left(c^{\dagger}_{j\sigma}c_{j+1,\sigma}+\mathrm{h.c.}\right)\nonumber\\
  &\quad +\frac{U}{2}(n_{j\uparrow}-1/2)(n_{j\downarrow}-1/2)\label{eq:local_bond_energy}\\
  &+\frac{U}{2}(n_{j+1,\uparrow}-1/2)(n_{j+1,\downarrow}-1/2),\nonumber
\end{align}
with the on-site interaction evenly distributed over adjacent bonds. The half-chain operator is then
\begin{equation}
E_{L/2}\equiv \sum_{j=1}^{L/2-1} h_j + \frac{1}{2}\,h_{L/2},
\label{eq:E_half_chain}
\end{equation}
where the boundary bond $h_{L/2}$ is split symmetrically between the two halves. The transferred energy variable is $\Gamma_E(t)\equiv E_{L/2}(t)-E_{L/2}(0)$, 
and the corresponding second moment defines the energy roughness,
\begin{equation}
\kappa_2^{(E)}(t)\equiv \big\langle\Gamma_E(t)^2\big\rangle.
\label{eq:kappa2_E_def}
\end{equation}
In the $U=0$ limit, the two spin species are independent, each governed by the tight-binding Hamiltonian~\eqref{eq:H0_sigma}, 
and the local bond energy operator 
reduces to the hopping term $\sim J$ in Eq. \eqref{eq:local_bond_energy}. The 
half-chain energy becomes a sum of two-body operators. Using the single-particle Heisenberg evolution from 
Eq.~\eqref{eq:U0_cj_evolution}, one can express the two-time
correlator of $E_{L/2}$ as a sum over products of single-particle propagators. For the infinite-temperature ensemble,
Wick's theorem again applies. A single spin species contributes
\begin{gather}
\langle h_j^{\sigma}(t)\,h_{j'}^{\sigma}\rangle
=\frac{J^2}{2}\left[J_{j-j'}(2Jt)^2-J_{j-j'-1}(2Jt)J_{j-j'+1}(2Jt)\right],
\label{eq:hh_corr_U0}
\end{gather}
Summing over all contributing bonds and using the same stationarity argument as in Sec.~\ref{sec:dynexp}
(i.e., $\langle E_{L/2}(t)^2\rangle=\langle E_{L/2}^2\rangle$), one obtains
\begin{equation}
\kappa_2^{(E)}(t) = 2\Big(\langle E_{L/2}^2\rangle - \langle E_{L/2}(t)\,E_{L/2}(0)\rangle\Big).
\label{eq:kappa2_E_stationarity}
\end{equation}
Writing $E_{L/2}=\sum_j w_j h_j$ with $w_j=1$ for $1\le j\le L/2-1$, $w_{L/2}=1/2$, and $w_j=0$ otherwise,
the overlap factor appearing in the bond sum can be evaluated explicitly. For $r\neq 0$ one finds
\begin{equation}
\sum_j w_j w_{j-r}=\left(\frac{L}{2}-\frac{3}{4}\right)-\left(|r|-\frac{1}{4}\right).
\label{eq:energy_weight_overlap}
\end{equation}
Using Eq.~\eqref{eq:hh_corr_U0} together with the identity
$\sum_r\left[J_r(x)^2-J_{r-1}(x)J_{r+1}(x)\right]=1$, the extensive piece cancels against
$\langle E_{L/2}^2\rangle$ in Eq.~\eqref{eq:kappa2_E_stationarity}. In the thermodynamic limit
($L\to\infty$, half-chain $\ell=L/2\to\infty$), one obtains
\begin{equation}
\kappa_{2}^{(E)}(t)=2J^2\sum_{r\neq 0}\left(|r|-\frac{1}{4}\right)
\left[J_r(2Jt)^2-J_{r-1}(2Jt)J_{r+1}(2Jt)\right].
\label{eq:kappa2_E_U0_exact}
\end{equation}
At short times $t\ll 1/J$, expanding the Bessel functions gives
\begin{equation}
\kappa_2^{(E)}(t)\simeq \frac{3}{2}J^4 t^2 + \mathcal{O}(t^4),\quad t\ll\frac{1}{J},
\label{eq:kappa2_E_short_time}
\end{equation}
reflecting the universal quadratic growth characteristic of all conserved quantities at the microscopic scale.
In the hydrodynamic regime $t\gg 1/J$, using $\sum_r r^2 J_r^2(2Jt)\simeq 2J^2t^2$ and the asymptotic decay
$J_r(2Jt)\sim (Jt)^{-1/2}$, the energy correlator grows linearly,
\begin{equation}
\kappa_2^{(E)}(t)\simeq c_E\, J^3 t,\quad t\gg\frac{1}{J},
\label{eq:kappa2_E_asymptotics}
\end{equation}
with a numerical prefactor $c_E$ of order unity. This linear growth $\kappa_2^{(E)}(t)\propto t$ indicates
\emph{ballistic} energy transport, implying $z=1$ in the energy sector for $U=0$, similar to the charge and
spin transport sectors. 
For the numerical QGF evaluation of the energy sector, we define the corresponding energy twist operator
\begin{equation}
R_{L/2}^{(E)}(\lambda)\equiv e^{i\lambda E_{L/2}}.
\label{eq:energy_twist_operator}
\end{equation}
Unlike the charge and spin twists in Eq.~\eqref{eq:twist_half_chain}, the energy twist does not factorize into a simple
tensor product of on-site operators because neighboring bond energies overlap and therefore do not commute,
$[h_j,h_{j+1}]\neq 0$. A convenient MPO construction is obtained by introducing
\begin{equation}
\tilde h_j\equiv
\begin{cases}
h_j, & 1\le j\le L/2-1,\\
\dfrac{1}{2}h_{L/2}, & j=L/2,\\
0, & j>L/2,
\end{cases}
\qquad E_{L/2}=\sum_{j=1}^{L-1}\tilde h_j,
\label{eq:energy_local_terms}
\end{equation}
and splitting the half-chain energy into odd and even bond contributions,
\begin{equation}
E_{L/2}^{\rm odd}=\sum_{j\,\mathrm{odd}} \tilde h_j,
\qquad
E_{L/2}^{\rm even}=\sum_{j\,\mathrm{even}} \tilde h_j.
\label{eq:energy_odd_even_split}
\end{equation}
Within each set the terms are mutually commuting, so for small counting field one may represent the twist operator by a
second-order Suzuki decomposition,
\begin{equation}
R_{L/2}^{(E)}(\lambda)
\approx
e^{i\lambda E_{L/2}^{\rm odd}/2}
e^{i\lambda E_{L/2}^{\rm even}}
e^{i\lambda E_{L/2}^{\rm odd}/2}
+\mathcal{O}(\lambda^3).
\label{eq:energy_twist_trotter}
\end{equation}
Equivalently,
\begin{align}
R_{L/2}^{(E)}(\lambda)
&\approx
\prod_{\substack{j\le L/2\\ j\,\mathrm{odd}}} e^{i\lambda \tilde h_j/2}
\prod_{\substack{j\le L/2\\ j\,\mathrm{even}}} e^{i\lambda \tilde h_j}
\prod_{\substack{j\le L/2\\ j\,\mathrm{odd}}} e^{i\lambda \tilde h_j/2},
\label{eq:energy_twist_layers}
\end{align}
which is a sequence of commuting two-site gates and can therefore be applied to the identity MPO exactly as in a TEBD
step, followed by the usual compression. This yields an efficient MPO representation of the energy twist operator both for real
$\lambda=r$ and for the imaginary phase $\lambda=i r$ 
entering Eq.~\eqref{eq:energy_twist_operator}.
\begin{figure}
\includegraphics[width=\columnwidth]{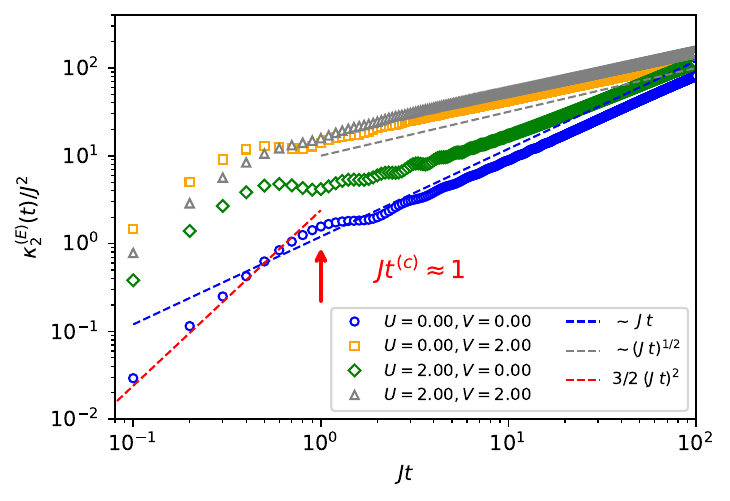}
\caption{Half-chain second moment $\kappa_2^{(E)}(t)$ as function of time. 
The dashed lines show the expected power-law growth with
exponents $1/z=1$ (ballistic) and $1/2$ (diffusive). 
The numerical data are obtained for $L=200$ sites and a small counting field $\lambda=0.05$.
The crossover time $t^{(c)}\sim 1/(J)$ is indicated by the vertical red arrow.}
\label{fig:kappa2_E_half_chain}
\end{figure}
For finite interactions ($U\neq 0$, $V=0$) the energy is no longer a simple bilinear and the Wick-theorem
derivation above no longer applies. Based on the GHD picture of the integrable Hubbard chain, energy transport at
infinite temperature is expected to be \emph{ballistic} ($z=1$) even in the interacting integrable case,
since the energy current does not couple to the non-Abelian charges that are responsible for the KPZ anomaly in the spin
and charge sectors. Breaking integrability ($V\neq 0$) is expected to convert the ballistic regime to
diffusive, $z=2$, in analogy with the charge and spin sectors. These expectations are summarized as
\begin{equation}
\kappa_2^{(E)}(t)\propto \begin{cases}
t^2, & t\lesssim t^{(c)},\\
 t, & t\gtrsim t^{(c)},\quad V=0 \text{ and } \forall U,\\
t^{1/2}, & t\gtrsim t^{(c)},\quad V\neq 0\text{ and } \forall U,
\end{cases}
\label{eq:kappa2_E_scaling_z}
\end{equation}
These expectations are presented in Fig.~\ref{fig:kappa2_E_half_chain}, where the TEBD data for the half-chain energy
second moment $\kappa_2^{(E)}(t)$ are shown together with dashed guides for the predicted asymptotic ballistic and
diffusive regimes, as well as the short-time crossover scale $t^{(c)}$.
These expectations parallel Eq.~\eqref{eq:kappa2_scaling_z} for the charge and spin sectors, but with the notable
absence of a KPZ scaling in the integrable interacting case. Within the accessible time window and system size, the TEBD
results are consistent with Eq.~\eqref{eq:kappa2_E_scaling_z}.

\section{Family--Vicsek scaling}
We now turn to the investigation of spatio-temporal fluctuation in a finite segment, providing both the growth and saturation exponents. This also establishes the FV universality in Hubbard chains.

\subsection{Scaling for charge and spin fluctuations}
\label{sec:fv_scaling}
%\subsection{Non-interacting limit \texorpdfstring{$U=0$}{U=0}}
\label{sec:U0_benchmark}
It is useful in the beginning to address the $U=0$ limit of the Hubbard model, which 
is analytically tractable and provides a benchmark for the QGF numerics. In this limit the model describes two independent species of free fermions, and the segment charge and spin operators can be expressed as
$N_\ell=N_\ell^{\uparrow}+N_\ell^{\downarrow}$ and
$S^z_\ell=\tfrac{1}{2}(N_\ell^{\uparrow}-N_\ell^{\downarrow})$, where
\begin{equation}
N_\ell^{\sigma}\equiv \sum_{j\in\mathrm{seg}(\ell)} n_{j\sigma}.
\end{equation}
We consider infinite-temperature ensembles which are diagonal in the occupation basis and translation invariant, so that
$\bar n_{\sigma}\equiv\langle n_{j\sigma}\rangle$ is independent of $j$. For the fully mixed state $\rho_{\infty}=\mathds{1}/4^L$,
one has $\bar n_{\uparrow}=\bar n_{\downarrow}=1/2$.
Defining the transferred variable for each spin species as
\begin{equation}
\Gamma_{\sigma}(t)\equiv N_\ell^{\sigma}(t)-N_\ell^{\sigma}(0),
\end{equation}
stationarity of the infinite-temperature ensemble implies
\begin{equation}
\langle\Gamma_{\sigma}(t)^2\rangle
=2\Big(\langle (N_\ell^{\sigma})^2\rangle-\langle N_\ell^{\sigma}(t)N_\ell^{\sigma}\rangle\Big).
\end{equation}
Evaluating $\langle N_\ell^{\sigma}(t)N_\ell^{\sigma}\rangle$ using Eq.~\eqref{eq:U0_nn_kernel} yields
\begin{align}
\langle\Gamma_{\sigma}(t)^2\rangle
&=\frac{1}{2}\left[\ell-\sum_{i,j\in\mathrm{seg}(\ell)}|U_{ij}(t)|^2\right]\\
&=\frac{1}{2}\left[\ell-\sum_{r=-(\ell-1)}^{\ell-1}(\ell-|r|)\,J_r^2(2Jt)\right],
\end{align}
where in the last step we used translation invariance of the kernel $U_{jm}(t)=i^{\,m-j}J_{m-j}(2Jt)$. Therefore,
\begin{equation}
\kappa_{2,\sigma}^{(0)}(\ell,t)
=\frac{1}{2}\left[\ell-\sum_{r=-(\ell-1)}^{\ell-1}(\ell-|r|)\,J_r^2(2Jt)\right],
\label{eq:mu2_U0}
\end{equation}
Because the two species are dynamically and statistically independent in this limit, the charge and spin moments follow as in
Eq.~\eqref{eq:kappa2_charge_spin_U0_exact}.
The corresponding roughness functions are $W_N(\ell,t)=\sqrt{\mu_{2,N}^{(0)}(\ell,t)}$ and
$W_{S^z}(\ell,t)=\sqrt{\mu_{2,S^z}^{(0)}(\ell,t)}$.
In the growth regime $1\ll 2Jt\ll \ell$, one may use the asymptotics
$\sum_{r=-\infty}^{\infty}|r|\,J_r^2(2Jt)\simeq \tfrac{4}{\pi}Jt$ to obtain
\begin{equation}
W_N(\ell,t)\simeq \left(\frac{4}{\pi}Jt\right)^{\!1/2}, \quad 
W_{S^z}(\ell,t)\simeq \frac{W_N(\ell,t)}{2}.
\end{equation}
At long times $t\gg \ell/(2J)$ the two-time correlators decorrelate and Eq.~\eqref{eq:mu2_U0} yields saturation to twice the
static variance,
\begin{equation}
W_{N,\mathrm{sat}}(\ell)\simeq \ell^{1/2},
\label{eq:W_N_U0_sat}\quad
W_{S^z,\mathrm{sat}}(\ell)\simeq \frac{W_{N,\mathrm{sat}}(\ell)}{2}.
\end{equation}
These results imply FV exponents $\beta=1/2$ and $\alpha=1/2$ (for both charge and spin), and hence a ballistic dynamical exponent
$z=\alpha/\beta=1$.
The crossover time scale can be estimated by matching the growth and saturation laws, giving
\begin{equation}
t^*(\ell)\simeq \frac{\pi}{4}\,\frac{\ell}{J}.
\label{eq:tstar_U0}
\end{equation}
Equivalently, using the scaling variable $x\equiv Jt/\ell$, one may write
$W_Q(\ell,t)=\ell^{1/2}f_Q(x)$ with sector-dependent amplitudes. For the charge sector, the asymptotics read
\begin{equation}
f_N(x)\simeq
\begin{cases}
\sqrt{\dfrac{4}{\pi}}\,x\,^{\!1/2}, & x\ll 1,\\
1, & x\gg 1,
\end{cases}
\end{equation}
while for the spin sector the prefactors differ by a 
factor of $2$, i.e., $f_{S^z}(x)=f_N(x)/2$.
\begin{figure}
\includegraphics[width=\columnwidth]{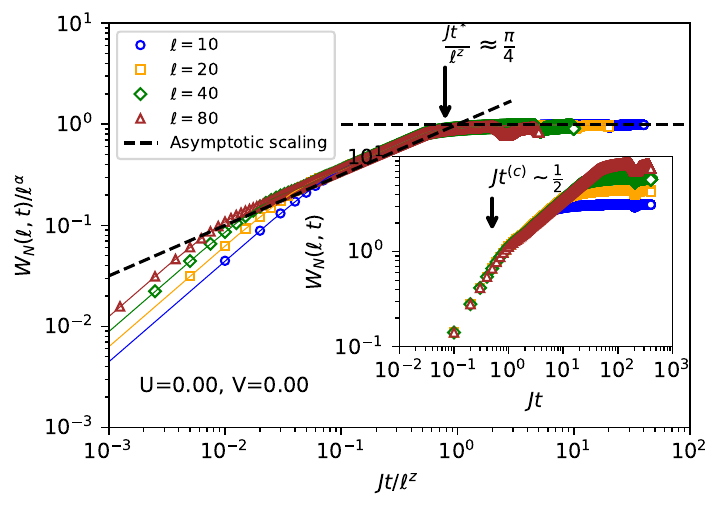}
\caption{Family--Vicsek scaling collapse for the charge roughness $W_N(\ell, t)$ for  $U=0$ and $V=0$. The data are obtained for $L=400$ sites and a small counting field $r=0.05$ in Eq.~\eqref{eq:mu2_phase_combo}. The dashed lines show the expected asymptotic power-law growth with $\beta=1/2$ and saturation with $\alpha=1/2$. The crossover time scale is consistent with the analytical estimate from Eq.~\eqref{eq:tstar_U0}. The inset shows the same data unrescaled, illustrating the three regimes separated by the crossover times $t^{(c)}$ and  $t^*(\ell)$.}
\label{fig:FV_scaling_U0_V0}
\end{figure}

Within the generating function approach, to construct the FV universal functions, it is necessary to examine the scaling of the second moments with respect to the subsystem size. This is achieved by
calculating the charge and spin fluctuations for varying subsystem size $l$, located in the middle of the chain. The initial twist operator is chosen as 
\begin{gather}
	R^{(Q)}(\lambda) =\mathbb{1}\otimes\dots \mathbb{1}\otimes\underbrace{e^{i\lambda Q_{\frac{L-l}{2}+1}}\otimes \dots \otimes e^{i\lambda Q_{\frac{L+l}{2}}}}_{\frac{L-l}{2}<j\le \frac{L+l}{2}}\otimes\mathbb{1}\otimes\dots \mathbb{1}.
  \label{eq:twist_operator_segment}
\end{gather}
Figure~\ref{fig:FV_scaling_U0_V0} demonstrates that the charge roughness in the free-fermion limit is fully 
consistent with the FV ansatz
$W_N(\ell,t)=\ell^{\alpha} f_N(Jt/\ell)$. Indeed, upon rescaling by $\ell^{1/2}$ (i.e., $\alpha=1/2$) the data for 
different segment lengths
collapse onto a single scaling curve. In the growth regime $Jt/\ell\ll 1$ the collapsed curve follows the expected 
power law
$W_N(\ell,t)\propto t^{\beta}$ with $\beta=1/2$, reflecting the linear growth of the second cumulant, $\kappa_2^{(N)}(t)\propto t$.
At late times $Jt/\ell\gg 1$, the roughness saturates at $W_{N,\mathrm{sat}}\propto \ell^{1/2}$, in agreement with 
Eq.~\eqref{eq:W_N_U0_sat}.
Matching these two asymptotic regimes yields the crossover scale $t^*(\ell)$, and the observed crossover is 
compatible with the analytical estimate
in Eq.~\eqref{eq:tstar_U0}. The inset (unrescaled data) highlights, in addition, the short-time microscopic regime 
separated by $t^{(c)}$ from the
hydrodynamic growth, followed by the FV crossover to saturation at $t^*(\ell)$. Although not displayed here, the 
spin roughness $W_{S^z}(\ell,t)$ shows the same scaling collapse with the same exponents, albeit with different 
prefactors.
%\subsection{Interacting integrable case \texorpdfstring{$U\neq 0$}{U!=0}, \texorpdfstring{$V=0$}{V=0}}
\begin{figure}
\includegraphics[width=\columnwidth]{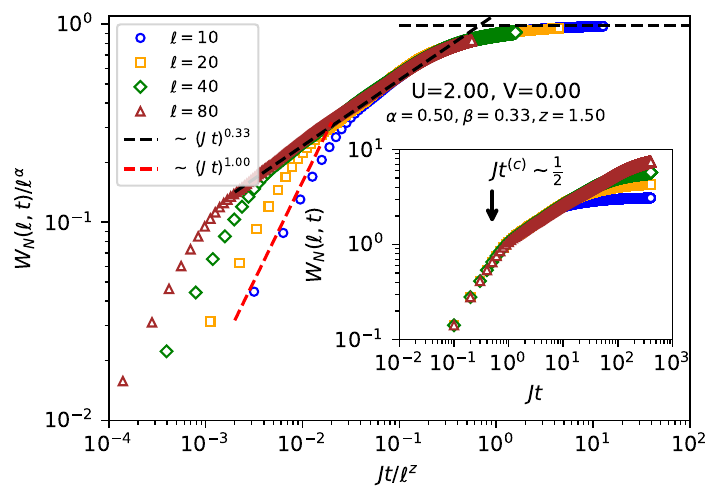}
\caption{Family--Vicsek scaling collapse for the charge roughness $W_N(\ell, t)$ for  $U=2.0$ and $V=0$, (interacting integrable model)
indicating the KPZ universality class, with dynamical exponent $z=3/2$, growth exponent $\beta=1/3$, and roughness exponent $\alpha=1/2$.	
 The black dashed lines show the expected asymptotic power-law growth with $\beta=1/3$. 
 The inset shows the same data unrescaled.}
\label{fig:FV_scaling_U2_V0}
\end{figure}
\begin{figure}
\includegraphics[width=\columnwidth]{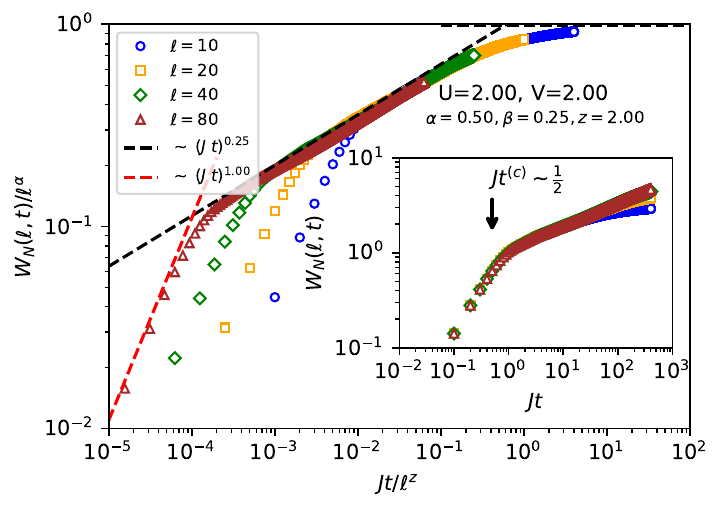}
\caption{Family--Vicsek scaling collapse for the charge roughness $W_N(\ell, t)$ for  $U=2$ and $V=2$, (non-integrable model) indicating a diffusive universality class, with dynamical exponent $z=2$, growth exponent $\beta=1/4$, and roughness exponent $\alpha=1/2$. The black dashed line indicates the expected asymptotic power-law growth with $\beta=1/4$, while the dashed red line indicates the 
expected asymptotic power-law growth in the short time microscopic regime with $\beta=1$ (ballistic).
The data are obtained for $L=400$ sites and a small counting field $r=0.05$ in Eq.~\eqref{eq:mu2_phase_combo}. The inset shows the same data unrescaled.}
\label{fig:FV_scaling_U2_V2}
\end{figure}

For finite interactions in the integrable Hubbard model ($U\neq 0$ and $V=0$), the FV analysis reveals a clear departure from the ballistic
free-fermions and a crossover to KPZ scaling. As shown in Fig.~\ref{fig:FV_scaling_U2_V0}, rescaling the charge roughness as
$W_N(\ell,t)/\ell^{\alpha}$ with $\alpha=1/2$ yields an excellent collapse of data for different subsystem sizes onto a universal curve as a function
of the scaling variable $t/\ell^{z}$. In the growth regime, the collapsed curve follows the KPZ power law
$W_N(\ell,t)\propto t^{\beta}$ with $\beta=1/3$, consistent with the half-chain result $\kappa_2^{(N)}(t)\propto t^{2/3}$ and implying the dynamical
exponent $z=\alpha/\beta=3/2$. At late times, the same data saturate as $W_{N,\mathrm{sat}}\propto \ell^{1/2}$, as expected for KPZ-type
roughening in one dimension. The inset illustrates the corresponding unscaled evolution and the onset of the KPZ growth window before saturation.

While our focus is on FV scaling of cumulants, it is useful to note that closely related KPZ signatures in integrable
many-body dynamics have been discussed from complementary perspectives in both theory and experiment~\cite{Ye.2022,Wei.2022,Bulchandani.2019,Zunkovic2013}.
%\subsection{Non-integrable case \texorpdfstring{$V\neq 0$}{V!=0}}

Breaking integrability by switching on the next-nearest-neighbor density interaction $V\neq 0$ drives the long-time dynamics into a diffusive
hydrodynamic regime. This is reflected in Fig.~\ref{fig:FV_scaling_U2_V2}, where the FV rescaling with $\alpha=1/2$ and $z=2$ collapses the
data for different subsystem sizes, and the growth regime exhibits the diffusive scaling $W_N(\ell,t)\propto t^{1/4}$ (equivalently,
$\kappa_2^{(N)}(t)\propto t^{1/2}$) before saturating to $W_{N,\mathrm{sat}}\propto \ell^{1/2}$. The dashed red guide emphasizes that, at very short times,
the dynamics remains in the microscopic ballistic regime with $W_N(\ell,t)\propto t$, followed by a crossover to the diffusive FV scaling window.
Within numerical accuracy, we find the same diffusive scaling structure in the spin sector.
\subsection{Energy sector}\label{sec:energy_fv_scaling}

\begin{figure*}
\includegraphics[width=\columnwidth]{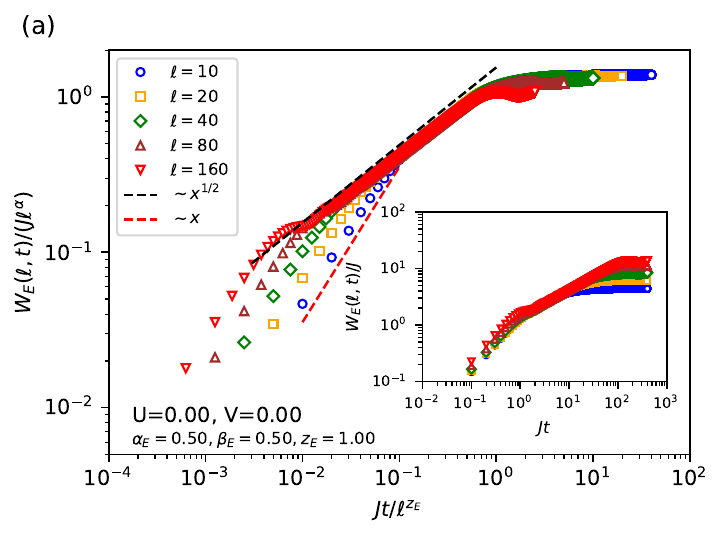}
\includegraphics[width=\columnwidth]{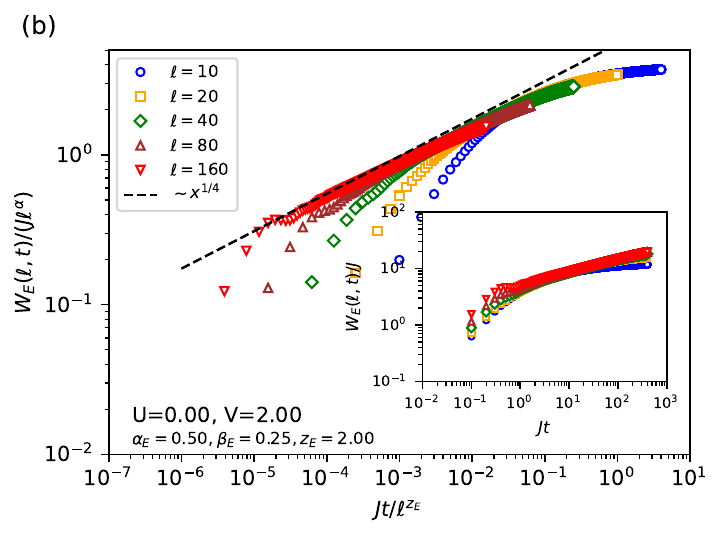}
\includegraphics[width=\columnwidth]{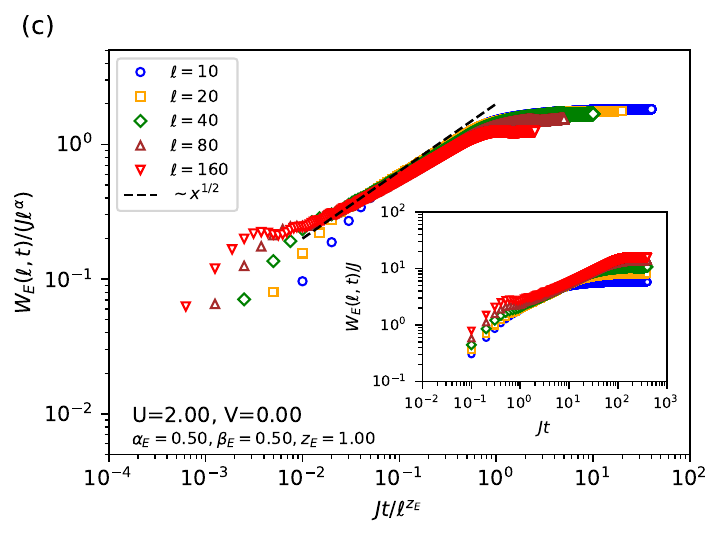}
\includegraphics[width=\columnwidth]{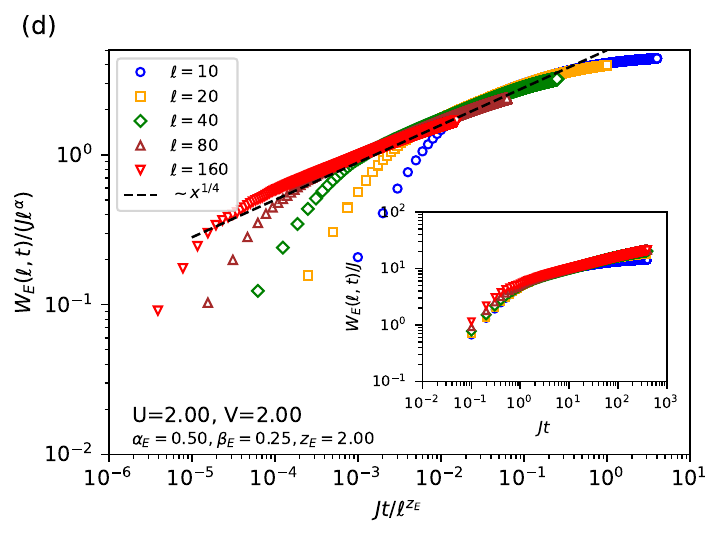}
\caption{Family--Vicsek scaling collapse for the energy roughness $W_E(\ell, t)$ for various values of $U$ and $V$. The black dashed line indicates the expected asymptotic power-law growth with $\beta_E=1/2$ (integrable) and $\beta_E=1/4$ (non-integrable), while the dashed red line in panel (a) indicates the 
expected asymptotic power-law growth in the short time microscopic regime.
The data are obtained for $L=400$ sites and a small counting field $r=0.05$ in Eq.~\eqref{eq:mu2_phase_combo}. The inset shows the same data unrescaled.}
\label{fig:FV_scaling_Energy}
\end{figure*}

To extend the Family--Vicsek analysis to energy transport, we define the energy roughness from the second cumulant of
the transferred energy in a segment of length $\ell$,
\begin{equation}
W_E(\ell,t)\equiv \sqrt{\kappa_2^{(E)}(\ell,t)}
\sim \ell^{\alpha_E}\, f_E\!\left(\frac{t}{\ell^{z_E}}\right),
\end{equation}
with $f_E(x\ll 1)\sim x^{\beta_E}$ and $z_E=\alpha_E/\beta_E$.

The collapse in Fig.~\ref{fig:FV_scaling_Energy} shows that the energy sector separates cleanly according to
integrability. In all four panels the late-time saturation is consistent with
$W_{E,\mathrm{sat}}(\ell)\propto \ell^{1/2}$, so the roughness exponent remains $\alpha_E=1/2$. The distinction appears
in the growth regime. For the integrable cases, panels (a) and (c) with $V=0$, the data collapse is obtained with the
scaling variable $t/\ell$ and the growth window follows $W_E(\ell,t)\propto t^{1/2}$, equivalently
$\kappa_2^{(E)}(\ell,t)\propto t$. This is the hallmark of ballistic energy transport, $z_E=1$. Panel (a),
corresponding to the analytically solvable $U=0$ limit, also resolves the universal short-time microscopic regime
$W_E\propto t$ before crossing over to the ballistic hydrodynamic window. The interacting integrable case in panel (c)
shows the same long-time collapse, indicating that, unlike the charge and spin sectors, the energy sector remains
ballistic when integrability is preserved.

By contrast, once integrability is broken, panels (b) and (d) collapse with $t/\ell^2$ and display the slower growth
$W_E(\ell,t)\propto t^{1/4}$, or $\kappa_2^{(E)}(\ell,t)\propto t^{1/2}$. This identifies the non-integrable energy
sector as diffusive with $z_E=2$. The agreement between the $U=0$, $V=2$ and $U=2$, $V=2$ panels shows that the
relevant distinction is not the mere presence of on-site interactions, but whether integrability is preserved. In other
words, Fig.~\ref{fig:FV_scaling_Energy} supports a simple difference: energy transport is ballistic in the integrable  limit, $V=0$, whereas the non-integrable cases with $V\neq 0$ are diffusive.

These observations are summarized by
\begin{equation}
(\alpha_E,\beta_E,z_E)=
\begin{cases}
\left(\frac{1}{2},\frac{1}{2},1\right), & V=0,\text{ and } \forall U\\
\left(\frac{1}{2},\frac{1}{4},2\right), & V\neq 0,\text{ and } \forall U
\end{cases}
\label{eq:energy_FV_exponents_interacting}
\end{equation}
in agreement with the exact free-fermion result and with the half-chain analysis of Sec.~\ref{sec:energy_dynexp}.

\begin{table}[h!]
\caption{Summary of Family--Vicsek scaling exponents for charge, spin, and energy sectors. The roughness exponent is $\alpha$, the growth exponent is $\beta$, and the dynamical exponent is $z=\alpha/\beta$. The transport regime is determined by the interaction parameters $U$ and $V$.}
\centering
\resizebox{\columnwidth}{!}{%
\begin{tabular}{|c|c|c|c|c|c|}
\hline
\textbf{Sector} & \textbf{Regime} & \textbf{Parameters} & $\alpha$ & $\beta$ & $z$ \\
\hline
\hline
\multirow{3}{*}{Charge/Spin} & Ballistic & $U=0, V=0$ & $1/2$ & $1/2$ & $1$ \\
\cline{2-6}
& KPZ & $U\neq 0, V=0$ & $1/2$ & $1/3$ & $3/2$ \\
\cline{2-6}
& Diffusive & $V\neq 0$ \text{and} $\forall U$ & $1/2$ & $1/4$ & $2$ \\
\hline
\hline
\multirow{2}{*}{Energy} & Ballistic & $V=0$ \text{and} $\forall U$ & $1/2$ & $1/2$ & $1$ \\
\cline{2-6}
& Diffusive & $V\neq 0$ \text{and} $\forall U$ & $1/2$ & $1/4$ & $2$ \\
\hline
\end{tabular}%
}
\label{tab:exponents}
\end{table}

We summarize the exponents in Table~\ref{tab:exponents}. 
 In the charge and spin sectors, the dynamics transitions from ballistic ($z=1$) in the free-fermion 
limit to KPZ-type superdiffusion ($z=3/2$) in the interacting integrable case, and finally to conventional 
diffusion ($z=2$) when integrability is broken. The energy sector, by contrast, remains ballistic in the entire 
integrable regime ($V=0$) and only becomes diffusive when integrability is explicitly broken. The roughness 
exponent $\alpha=1/2$ is universal across all sectors and regimes, consistent with expectations for one-dimensional 
systems.

\section{Conclusions}
\label{sec:conclusions}

In this work, we used a quantum generating function approach together with Family--Vicsek scaling to study how charge, spin, and energy fluctuations spread in the one-dimensional Hubbard model at infinite temperature. 
This framework allows us to compare different subsystem sizes on the same footing and identify the transport regime directly from the scaling collapse of the roughness.

The central conclusion is that integrability controls the long-time dynamics. In the free limit, all three sectors show ballistic behavior. 
In the interacting but still integrable Hubbard chain, charge and spin dynamics transition to  KPZ regime, whereas the energy sector remains ballistic. Once integrability is broken by the next-nearest-neighbor density interaction, the long-time behavior becomes diffusive in charge, spin, and energy alike.

We also find a common short-time microscopic regime before the hydrodynamic scaling window emerges. The later-time behavior, however, clearly separates the different universality classes and shows that energy follows a different route from charge and spin along the integrable line.

More broadly, these results show that Family--Vicsek scaling provides a compact way to organize ballistic, anomalous, and diffusive transport within a single framework for the Hubbard model. 
A natural next step would be to explore how this picture changes when the symmetry structure of the model is modified while integrability is still preserved.

\begin{acknowledgments}
This work received financial support from CNCS/CCCDI-UEFISCDI, under 
projects number PN-IV-P1-PCE-2023-0159 and PN-IV-P1-PCE-2023-0987 and by the National Research, Development and Innovation Office - NKFIH  Project No. K142179.
We acknowledge the Digital Government Development and Project
Management Ltd.~for awarding us access to the Komondor HPC facility based in Hungary. We acknowledge  
the use of the computing infrastructure provided by the University of Oradea and by IOSIN-PACTES at the Institute of Space Science - INFLPR Subsidiary, Bucharest-Magurele.
\end{acknowledgments}

\bibliography{references}

\end{document}